\documentclass[journal]{IEEEtran}
\usepackage{xcolor,soul,framed} %,caption
\colorlet{}{yellow}
\usepackage{graphicx}
\usepackage[cmex10]{amsmath}
\usepackage{array}
\usepackage{cite}
\usepackage{mdwmath}
\usepackage{mdwtab}
\usepackage{eqparbox}
\usepackage{url}
\usepackage{amsfonts, amssymb}
\usepackage{subfigure}
\usepackage{epstopdf}
\usepackage{wrapfig}
\begin{document}
\bstctlcite{}
    \title{Regions of Attraction Estimation using Level Set Method for Complex Network System}
  \author{Mengbang Zou, Yu Huang,
      Weisi Guo$^*$,~\IEEEmembership{Senior Member,~IEEE,}\\

\thanks{M. Zou is with Cranfield University, Cranfield, MK43 0AL, U.K.}
\thanks{Y. Huang is with South China University of Technology, Guangzhou 510641, China, and also with Cranfield University, MK43 0AL, U.K.}
\thanks{W. Guo (corresponding author)is with Cranfield University, Cranfield MK43 0AL, U.K., and also with the Alan Turing
Institute, London, NW1 2DB, U.K. (e-mail: weisi.guo@cranfield.ac.uk).} }

\maketitle

\begin{abstract}
Many complex engineering systems network together functional elements and balance demand loads (e.g. information on data networks, electric power on grids). This allows load spikes to be shifted and avoid a local overload. In mobile wireless networks, base stations (BSs) receive data demand and shift high loads to neighbouring BSs to avoid outage. The stability of cascade load balancing is important because unstable networks can cause high inefficiency. The research challenge is to prove the stability conditions for any arbitrarily large, complex, and dynamic network topology, and for any balancing dynamic function.

Our previous work has proven the conditions for stability for stationary networks near equilibrium for any load balancing dynamic and topology. Most current analyses in dynamic complex networks linearize the system around the fixed equilibrium solutions. This approach is insufficient for dynamic networks with changing equilibrium and estimating the Region of Attraction (ROA) is needed. The novelty of this paper is that we compress this high-dimensional system and use Level Set Methods (LSM) to estimate the ROA. Our results show how we can control the ROA via network topology (local degree control) as a way to configure the mobility of transceivers to ensure preservation of stable load balancing.
\end{abstract}


\IEEEpeerreviewmaketitle

\section{Introduction}

\IEEEPARstart{E}{ngineering} complex systems often comprise of a network of functional nodes (e.g., transceivers, pumps, capacitors, amplifiers), connected via edges (e.g., data links, pipes, electric lines, optic fibres). Together they serve multiple end-users' demand (e.g., multiple access communication, safe drinking water, synchronized electric power). Often there is a local demand surge, which can cause one node to be overwhelmed by demand and load balancing to nearby nodes is required. For example, in wireless communication networks, load balancing between adjacent base stations (BSs) \cite{8931743}, \cite{hu2019intelligent} is possible either via handing off edge users or coverage expansion \cite{6502480}. 

\subsection{Review on the Stability of Complex Network Systems}

Cascade load balancing arises when a BS passes load to a neighbour which in turn becomes overloaded and need to pass some of its load to the next neighbour, etc. This doesn't necessarily reduce overall network outage and can lead to significant inefficiencies due to information passing. Against this background, endless load balancing should be avoided in wireless communication network. 

\subsubsection{Review on Near Equilibrium Methods}

Recent breakthroughs have developed the framework to analysis the relationship between microscopic (like local component dynamics), macroscopic (global dynamics) behaviors and network topology \cite{gao2016universal,9268097,moutsinas2020node,zou2020uncertainty}. Our previous work in \cite{8931743} proved that provided 3 conditions are met: any network of arbitrarily large size, topological complexity, and balancing function form is stable. The conditions are: (1) the network topology is static, (2) the state of the network is not far from equilibrium (related to 1), and (3) knowledge of network parameters has a smaller variance than any noise in the system. 

\subsubsection{Review on Region of Attraction for Non-Equilibrium Problems}

Now, for example, if we consider a dynamic network where BSs can move (e.g., that of a UAV communication network \cite{jabbar2017energy}), clearly conditions (1) and (2) are violated. So the research question is, to what extent load changes or network topology changes will break the stability of load balancing? Estimating of the region of attraction (ROA) is needed when dealing with this problem. The ROA is the set of initial states from which the system converges to the equilibrium point \cite{wu2013exact}. Studies on ROA of load balancing in wireless network can identify how to maintain or control the stability of the network. There are some studies on the stability of load balancing in wireless network \cite{8931743, sun2019location}. Yet, to the best of our knowledge, no one has paid attention to estimate the ROA for load balancing on dynamic wireless network before, especially considering the topology change of large-scale networks.

The problem of estimating ROA has attracted a lot of attention in various fields, like mechanical systems, biological system, power systems and so on \cite{anghel2013algorithmic, matthews2012region, chakraborty2011nonlinear}. The challenge is that finding the exact ROA of a nonlinear system is a very difficult problem, because of the complex dynamics of nonlinear systems \cite{AWREJCEWICZ20211143}. In a large-scale wireless network, a large number of BSs couple together and affect each others' dynamics, which is typically a high-dimensional nonlinear complex system. Even though some methods have been successfully proposed to estimate ROA  \cite{AWREJCEWICZ20211143, yuan2019estimation, matallana2010estimation}, it is impossible to directly use these methods to estimate the ROA of such a high-dimensional nonlinear complex network system. Besides, we do not understand the relationship between network topology and local dynamics of each BSs in such a networked system.

\subsection{Novelty}

In this paper, we propose an analysis framework to estimate the ROA for traffic load balancing on wireless network. The key step is that we compress the high-dimensional dynamics to low-dimensional from the perspective of network topology. Based on this, we can use the Level Set Methods (LSM) \cite{yuan2019estimation} to estimate the ROA of this complex network system. Besides, we analyse the relationship between the local dynamics of each BS and network topology (local degree), which can identify how network topology affects the ROA of BSs. This study will give us insight into how to deployment BSs in wireless network from the perspective of network topology to enlarge the ROA, which means that the system can be more stable under perturbation. This study also identify how to control the change of traffic load to maintain the stability of BSs.

The remainder of this paper is organised as follows. Section \uppercase\expandafter{\romannumeral2} builds the dynamics model for load balancing in complex wireless network. In section \uppercase\expandafter{\romannumeral3}, we present the analytical framework to estimate the ROA of this system. In section \uppercase\expandafter{\romannumeral4}, we show the results of estimation of ROA and analyse the relationship between network topology and the ROA. Finally, the conclusion of our work is drawn and the future direction is foreseen in section \uppercase\expandafter{\romannumeral5}.

\section{System Setup}
In this section, we will build the dynamics model for load balancing of BSs in complex wireless network. We assume that there are $N$ BSs in the wireless network and two time scales: long-term traffic variations (traffic variation time scale $T$, in seconds) and short-term load dynamics under some constant traffic demand (symbol period time scale $t$, in milliseconds). In this paper, the short-term load dynamics is mainly concerned. Each BS $i$ has a load defined by $l_i(t) = d_i(T)/c_i(t)$, which is the ratio between the quasi-static long-term traffic demand aggregated across all users $u$ in BS $i$, $d_i(T) = \sum_ud_{i,u}(T)$, and the BS aggregated area capacity over all all users $u$ in BS $i$, $c_i(t) = \sum_uc_{i, u}(t)$. It is assumed that the capacity of each BS is stationary, which means that the capacity will not be dramatically affected by the change of load balancing.

\begin{figure}[tb]
\centering
\resizebox*{6cm}{!}
{\includegraphics{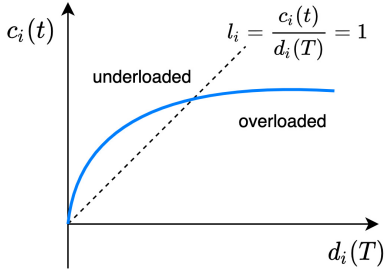}}
\caption{It shows the capacity $c_i(t)$ reacts to the traffic demand $d_i(T)$. At first, BS $i$ is underloaded. With the increase of traffic demand, the capacity of BS $i$ increases to meet the demand. When $c_i(t)$ saturate and $l_i>1$, BS $i$ need to offload traffic to neighbour BSs to avoid the outrage.} 
\label{fig_node_laod_dynamics}
\end{figure}

\begin{figure}
    \centering
    \resizebox*{9cm}{!}
    {\includegraphics{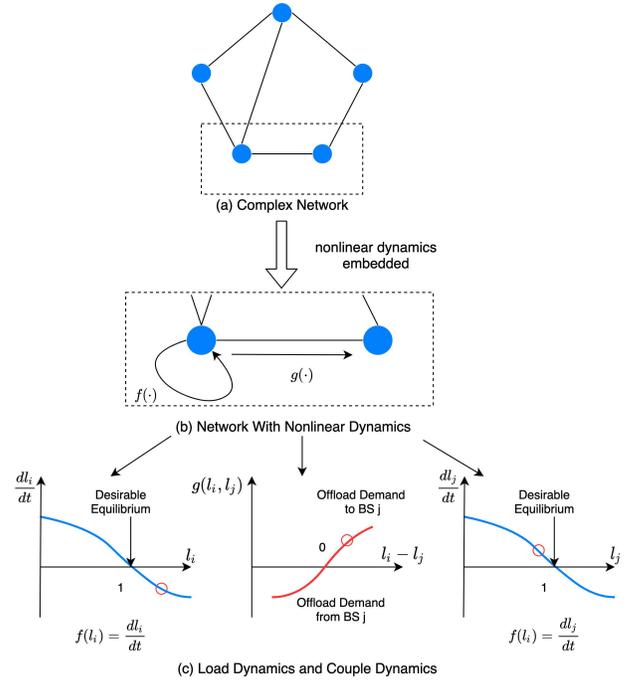}}
    \caption{Complex Wireless Network with Nonlinear Dynamics. (a) The wireless network contains a large number of BSs represented by nodes. All nodes are coupled together forming the complex network. (b) shows the coupled dynamics of node $i$ in complex network. The dynamics of BS $i$ is a linear combination of self dynamics $f(l_i)$ and coupling dynamics $g(l_i-l_j))$. (c) shows the load dynamics of BS $i, j$ $f(l_i), f(l_j)$ and the coupling dynamics $g(l_i,l_j)$. The left one characterizes the self load control of BS $i$. The middle one shows the coupling dynamics between BS $i$ and $j$. The right one shows the dynamics of BS $j$. When BS $i$ is overloaded, it will offload to the neighbour BS $j$ with light load to avoid outrage.}
    \label{fig_network_dynamics}
\end{figure}

For the single BS, the capacity $c_i(t)$ reacts to the traffic demand $d_i(T)$ (shown in Fig.~\ref{fig_node_laod_dynamics}). The self control of each BS tends to drive the load to a desirable equilibrium. When the capacity of BS $i$ saturates, BS $i$ needs to offload traffic load to neighbour BSs to avoid outrage. Therefore, there exists coupling dynamics between BSs. The overall network load balancing system dynamics is the linear combination of the self dynamics and coupling dynamics (shown in Fig.~\ref{fig_network_dynamics}). They are coupled together via the complex network, characterized by the weight matrix $a_{ij}$. The networked dynamics model has three main characteristics: (1) The number of nodes in the network is arbitrarily large. (2) The network is arbitrarily complex. (3) All of the nodes in the network are nonlinear dynamics embedded. The dynamics of BS $i$ in the network is described by

\begin{equation}\label{equ1}
    \dot{l_i}=f(l_i) + \sum^N_{j=1}a_{ji}g(l_j-l_i),
\end{equation}
where $a_{ji} = A_{ji}$ and $A$ is the connectivity matrix of the network. In this system, $l_i=1$ is a desirable equilibrium. So, what we are interested in is to estimate the ROA of this equilibrium. In this paper, we mainly consider two kinds of unstable behaviors as follows. First, the introduction of users within the BS leads to the variation of its traffic load $l_i$. Second, the mobility of BSs (e.g. Unmanned Aerial Vehicles (UAVs) supported base station) causes the change of network topology $a_{ij}$.

\section{Method to Estimate ROA}
\begin{figure}
    \centering
    \resizebox*{9cm}{!}
    {\includegraphics{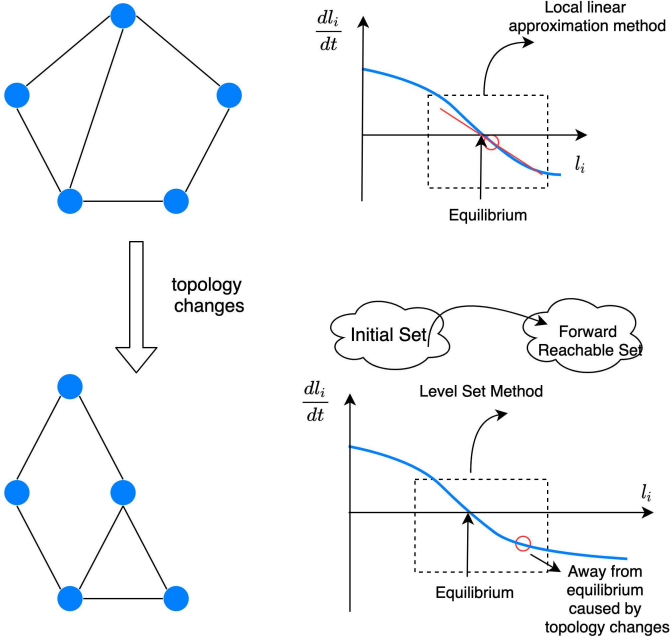}}
    \caption{In stationary complex network, it is usually to linearize the system around the equilibrium. When the network topology changes, the node may move away from the equilibrium. We need to estimate the ROA to know what extent perturbation will break the stability.}
    \label{fig_method}
\end{figure}

In a stationary complex network, we usually linearize the system near the equilibrium to analyze the stability. However, in a dynamic complex network, the change of network topology will cause the state of node far away from the equilibrium (shown in Fig.~\ref{fig_method}). As a result, it is not sufficient to linearize such system around the equilibrium. It is more important to estimate the ROA. Here we propose a analysis framework to estimate ROA in complex wireless network system with dynamics.  

\subsection{ROA for Linear System}
Firstly, we analyse the simple and ideal linear system. The load dynamics in BS $i$ can be described as
\begin{equation}
f(l_i)= \beta(1-l_i), 
\end{equation}
where $\beta>0$. If $\beta<0$, then the BS $i$ attempts to continue attract load demand from other BSs when BS $i$ is overloaded. When BS $i$ is underloaded, it still offload to neighbour BSs. This scenario is not common in most wireless communication systems. As a result, we do not consider this situation in this paper. $g(x)$ can be written as the Taylor expansion
\begin{equation}\label{equ2}
    g(x) = \sum_{n=0}^N\frac{f^{(n)}(a)}{n!}(x-a)^n+O[(x-a)^n],
\end{equation}
where $g(x)=0$,
and only consider the first order of $g(x)$. So $g(x) = \gamma x+ O(x^2)$, where $\gamma>0$ and $O(\cdot)$ is the big $O$ notation which bounds asymptotic behavior at $x=0$. If $\gamma<0$, BS $i$ attempts to share load demand to neighbour BSs, even though the load demand of BS $i$ is less than other BSs, which happens when BS $i$ is in sleep mode. However, such situation is not considered in this paper based on the assumption that all BSs are in active mode. Accordingly, we do not consider this situation. $\dot{l_i}$ can be characterised as
\begin{equation}
    \dot{l_i}=\beta(1-l_i) + \sum_{j=1}^Na_{ji}\gamma(l_j-l_i).
\end{equation}
For linear system, $l_i = 1$ is the only equilibrium. Then the stability of the equilibrium can be determined according to the eigenvalue of the Jacobian of the system \cite{8931743} as
\begin{equation}
    J(l_i=1)=\beta I-\gamma D+\gamma A^{T}= -\beta I -\gamma \Lambda^{T},
\end{equation}
where $J$ represents the Jacobian  of the system, $I$ is the identity matrix, $D$ is the weighted in-degree matrix, $\Lambda$ represents the weighted in-Laplacian of the graph and $\Lambda ^T$ denotes the transpose operation. Since $\beta > 0, \gamma>0$, $J(l_i=1)<0$. Therefore, the equilibrium $l_i=1$ is asymptotically stable. For a linear system, this equilibrium is asymptotically stable in large, and the ROA is the whole space.

\subsection{ROA for Nonlinear System}
In a nonlinear system, there may exist several equilibria, existing methods fail to directly estimate the ROA of each equilibrium due to the high dimension of the system. Here we compress the high dimension network to a low dimension by mapping the overall effective dynamics of a networked system to its topological structure and individual dynamics. It is given by
\begin{equation}
    \dot{l_i} = f(l_i) + w_ig(\bar{l}-l_i)
\end{equation}
and
\begin{equation}
    \dot{\bar{l}} = f(\bar{l}),
\end{equation}
where $w_i = \sum_{j=1}^Na_{ji}$ and $\bar{l}$ is the average load of the system. This estimation method is suitable for homogeneous network but not for heterogeneous network. By doing this, we compress the high dimensional system to a two-dimensional system, which can be estimated by the LSM.

To use the LSM to estimate the ROA, firstly, we need to prove the relationship between the exact ROA and the backward reachable set. Consider the following nonlinear system 
\begin{equation}\label{equ4}
    \dot{x}=f(x),
\end{equation}
where $f$ is Lipschitz continuous and $x_e$ is one equilibrium point of the system which satisfies $f(x_e)=0$. Let $\zeta_f(t;x_0,t_0)$ be the solution of equation (\ref{equ4}) starting at the initial state $x_0$ at initial time $t=0$. The region of attraction is defined as
\begin{equation}\label{equ5}
    R_f(x_e) := \left\{x\in \mathbb{R}|\lim\limits_{t \to \infty}\zeta_f(t;x,0)=x_e\right\}.
\end{equation}

Here we introduce two important concepts, forward reachable set and backward reachable set. The forward reachable set is defined as the set of all states which can be reached along trajectories within a certain time $t$ starting from the specified initial set $I$\cite{jin2005power}. The following function can describe the forward reachable set
\begin{equation}
\begin{split}
     F_f(I,[0,t]):=\\
      \{x\in \mathbb{R}|&
      \exists x_0 \in I,
      \exists t_f \in [0, t], \zeta_f(t_f; x_0, 0)=x \}.
\end{split}
\end{equation}

\begin{figure}
    \centering
    \resizebox*{6cm}{!}
    {\includegraphics{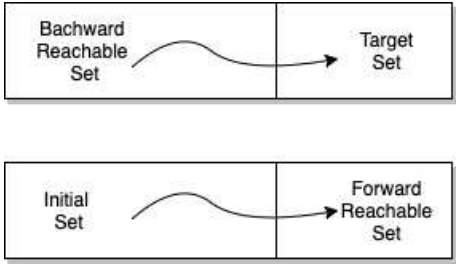}}
    \caption{Demonstration of the backward reachable set and forward reachable set.}
    \label{fig1}
\end{figure}

The backward reachable set is defined as the set of states where trajectories can reach the specified target set $K$ within time $t$. It is given by
\begin{equation}
\begin{split}
    B_f(S,[-t, 0]) := \\
    \{x_0\in \mathbb{R}|\exists x \in K, 
    &\exists t_f \in [-t, 0], \zeta_f(t_f;x_0,-t)=x\}. 
\end{split}
\end{equation}
The forward reachable set and backward reachable set are shown in in Fig.~\ref{fig1}.
The relationship between backward reachable set and ROA has been proved in \cite{yuan2019estimation}. The following function describes the connection
\begin{equation}
    A_f(x_e)=B_f(\Omega, (-\infty, 0]),
\end{equation}
where $x_e \in \Omega ^ {\circ}$, $\Omega \subset A_f(x_e)$.
Therefore, the exact ROA can be determined by computing the backward reachable set.

Following this strategy, the next step is to approximate the backward reachable set. Some methods based on Hamilton-Jacobi equation have been proposed to estimate the backward reachable set. Consider the backward reachable set from a target set $K_0 = \{x \in \mathbb{R}|\Xi(x) \le 0\}$ for a nonlinear system with dynamics
\begin{equation}
    \dot{x}=h(x, a, b),
\end{equation}
where $x$ is the state variable, $a(\cdot)$ is the input steering the system away from the target and $b(\cdot)$ is an input steering the system into target, $h:\mathbb{R}^n*M*N\xrightarrow[]{}\mathbb{R}^n$. Let $K$ be the set of states from which the trajectories will enter the target set $K_0$ within time $\tau$, then $K(\tau)=B_h(K_0, [-\tau, 0])$. $K_\tau$ can be determined by solving for the viscosity solution of a time-dependent Hamilton-Jacobi equation \cite{mitchell2005time}. Let $v:\mathbb{R}^n*[-T, 0]\xrightarrow[]{}\mathbb{R}$ be the viscosity solution of the terminal value Hamilton-Jacobi equation
\begin{equation}
\begin{split}
    \frac{\partial v(x, t)}{\partial t}+min[0, H(x, \frac{\partial v(x, t)}{\partial x})]=0\\
    v(x, 0) = \Xi(x),
\end{split}
\label{equ14}
\end{equation}
where 
\begin{equation}
    H(x, p) = \mathop{max}_{a \in M}\mathop{min}_{b \in N}p^Th(x, a, b).
    \label{equ15}
\end{equation}
Since the networked system in this paper does not have input, it is a special case of equation (\ref{equ14}), which can be directly solved by equation (\ref{equ14}). Therefore, the Hamilton-Jacobi equation can be written as

\begin{equation}
     \frac{\partial v(x, t)}{\partial t}+min[0, \frac{\partial v(x, t)}{\partial x}^Th(x)]=0\\
\end{equation}
Then 
\begin{equation}
    K(\tau) = \{x \in \mathbb{R}^n|v(x, -\tau)\le 0\}.
\end{equation}
So we can get 
\begin{equation}
    B_h(K_0, [-\tau, 0])= S_0(v(x, -\tau)),
\end{equation}
where $s_0(v)$ is the zero-sublevel set, $S_0(v):=\{x\in \mathbb{R}^n|v(x, t) \le 0\}$. Then, the ROA can be determined by $v(x, t)$ as
\begin{equation}
    A_f(x_e) = \{x \in \mathbb{R}^n|v(x, -\infty)\}.
\end{equation}
Since $min[0, H(x, \frac{\partial v(x, t)}{\partial x})] \le 0$, $\frac{\partial v(x, t)}{\partial t} \ge 0$. For $0<T<\infty, v(x, -T) \ge v(x, -\infty)$. Therefore, $S_0(v(x, -T))\subset \{x \in\mathbb{R}^n|v(x, -\infty)\le 0 \}$. This means that a conservative estimation of ROA can be approximated by $S_0(v(x, -T))$. So, an important step of this method for approximating ROA is to compute the viscosity solution of Hamilton-Jacobi equation. 
Here we use weighted essentially non-oscillatory (WENO) schemes to estimate $\frac{\partial v}{\partial x}$ \cite{jiang2000weighted}.

Let $x_k$ be a discretization of $R^1$ with uniform spacing $\Delta x$ and set
\begin{equation}
    v_k = v(x_k),  \Delta^+v_k=v_{k+1}-v_k, \Delta^-v_k-v_{k-1}.
\end{equation}
To estimate $v_x(x_i)$ on a left-biased stencil $\{x_k, k=i-3,...,i+2\}$, the $3^{rd}$ order accurate essentially non-oscillatory (ENO) will choose one from the following
\begin{equation}
\begin{split}
      & v_{x,i}^{-,0} = \frac{1}{3} \frac{\Delta v_{i-3}}{\Delta x} - \frac{7}{6} \frac{\Delta^+v_{i-2}}{\Delta x} + \frac{11}{6} \frac{\Delta^+v_{i-1}}{\Delta x}\\
      & v_{x,i}^{-,1} = -\frac{1}{6} \frac{\Delta v_{i-2}}{\Delta x} + \frac{5}{6} \frac{\Delta^+v_{i-1}}{\Delta x} + \frac{1}{3} \frac{\Delta^+v_{i}}{\Delta x} \\
      & v_{x,i}^{-,2} = \frac{1}{3} \frac{\Delta v_{i-1}}{\Delta x} + \frac{5}{6} \frac{\Delta^+v_{i}}{\Delta x} - \frac{1}{6} \frac{\Delta^+v_{i+1}}{\Delta x},
\end{split}    
\end{equation}
where $v_{x, i}^{-, s}$ is the $3^{rd}$ order approximation to $v_x(x_i)$ based on the $s^{th}$ substencil ${x_k, k=i+s-3,...,i+s}$ for $s=0,1,2$. The WENO approximation of $v_x(x_i)$ is a convex combination or weighted average of $v_{x, i}^{-,s} (s=0,1,2)$. Here, $w_s\le 0$ is the weight associated with the $s^{th}$ substencil and the weights satisfy the consistency equality: $w_0 + w_1 + w_2 = 1$. 
\begin{figure*}[tb]
\centering
\subfigure[ROA of system when $w_i=2$]{%
\resizebox*{6cm}{!}{\includegraphics{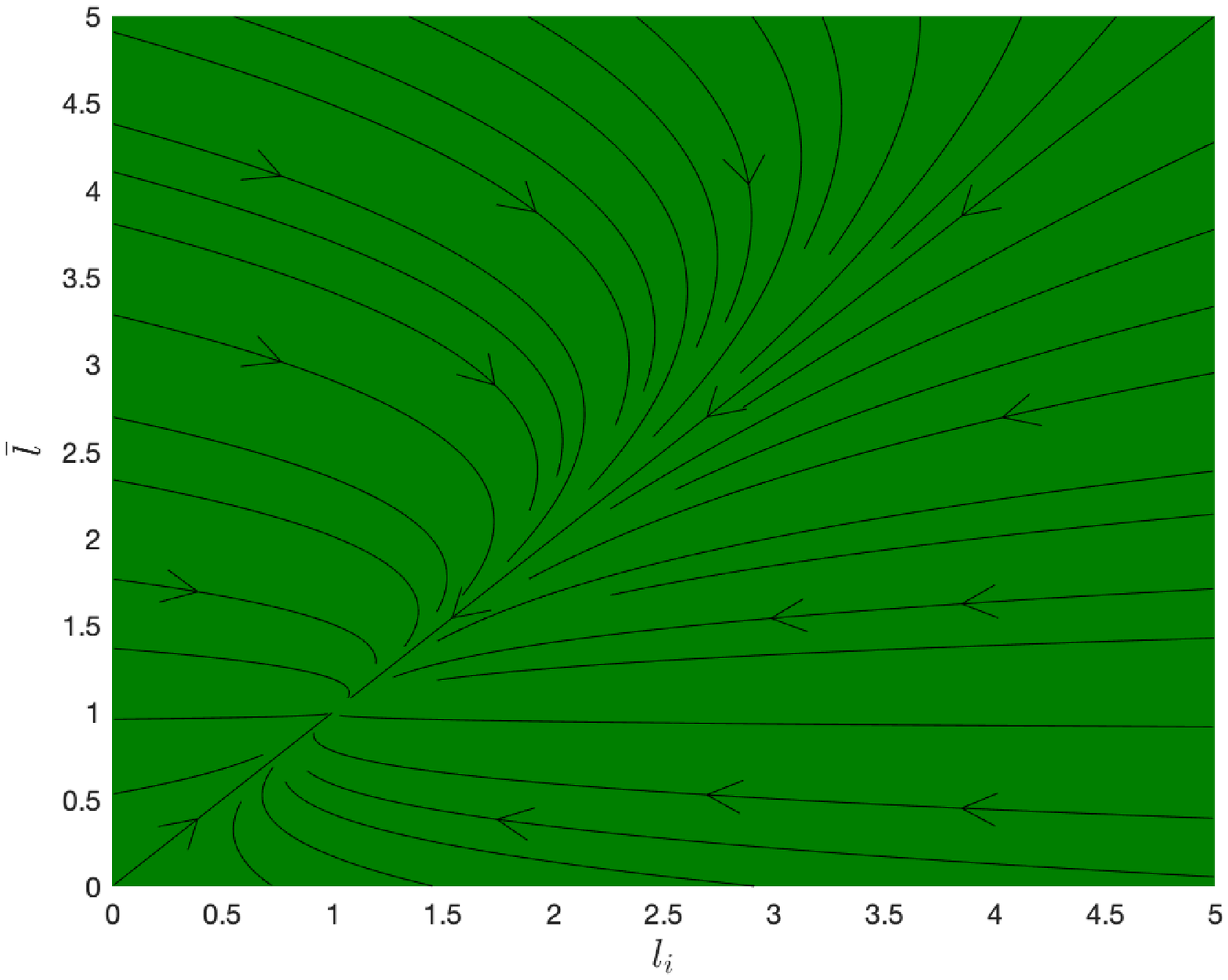}}}\hspace{2pt}
\subfigure[ROA of system when$w_i=4$]{%
\resizebox*{6cm}{!}{\includegraphics{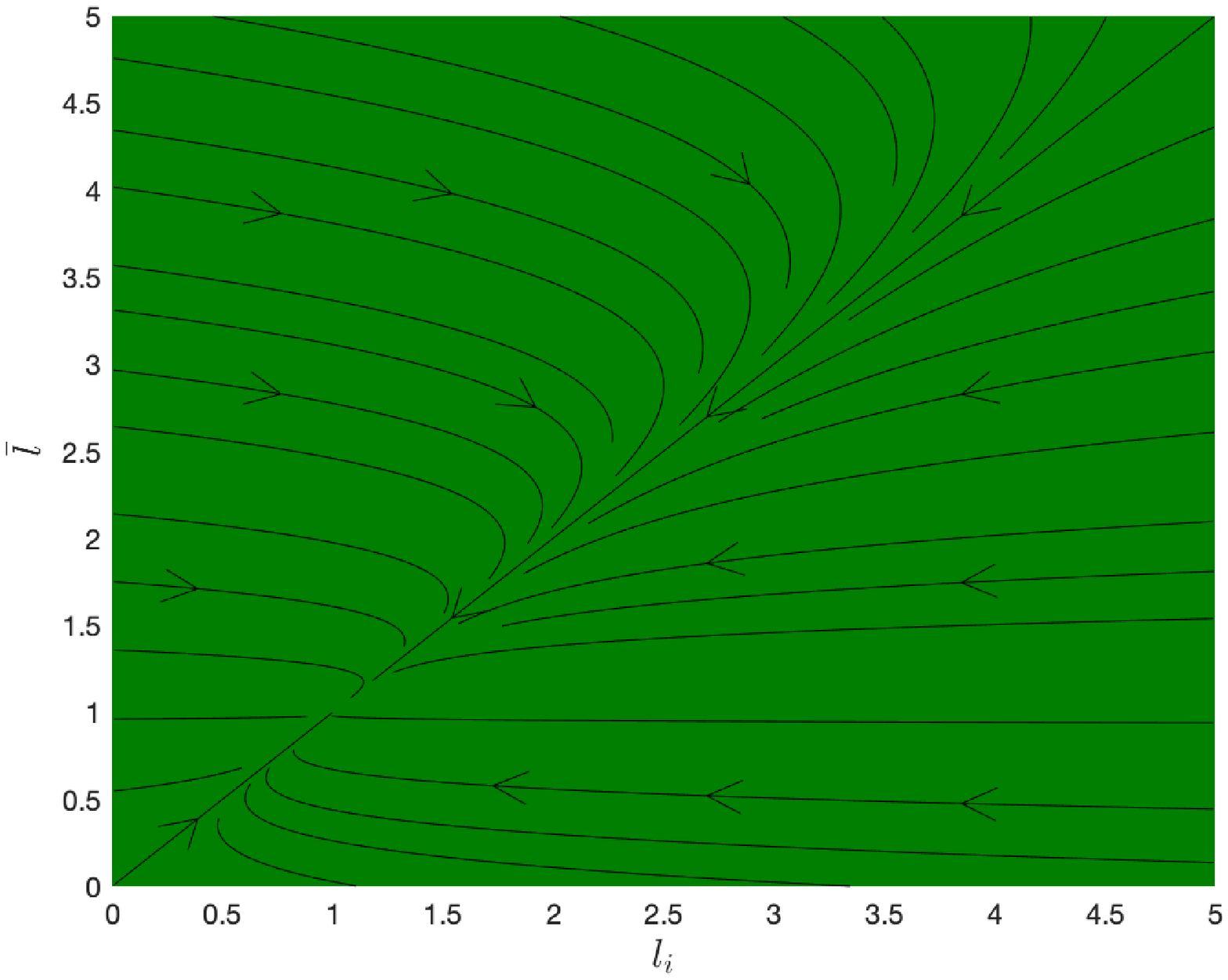}}}\hspace{2pt}
\subfigure[ROA of system when$w_i=6$]{%
\resizebox*{6cm}{!}{\includegraphics{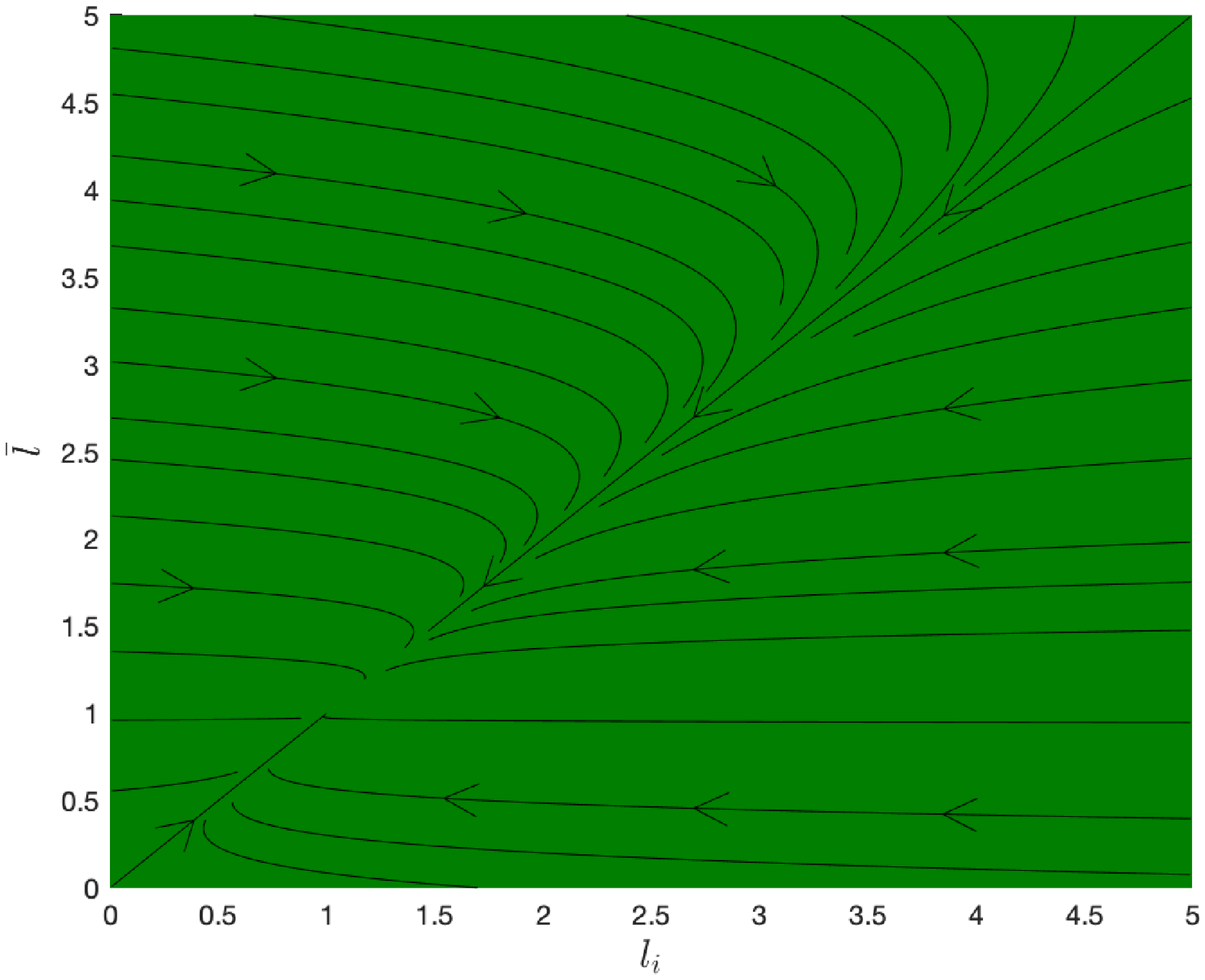}}}\hspace{2pt}
\subfigure[ROA of system when$w_i=8$]{%
\resizebox*{6cm}{!}{\includegraphics{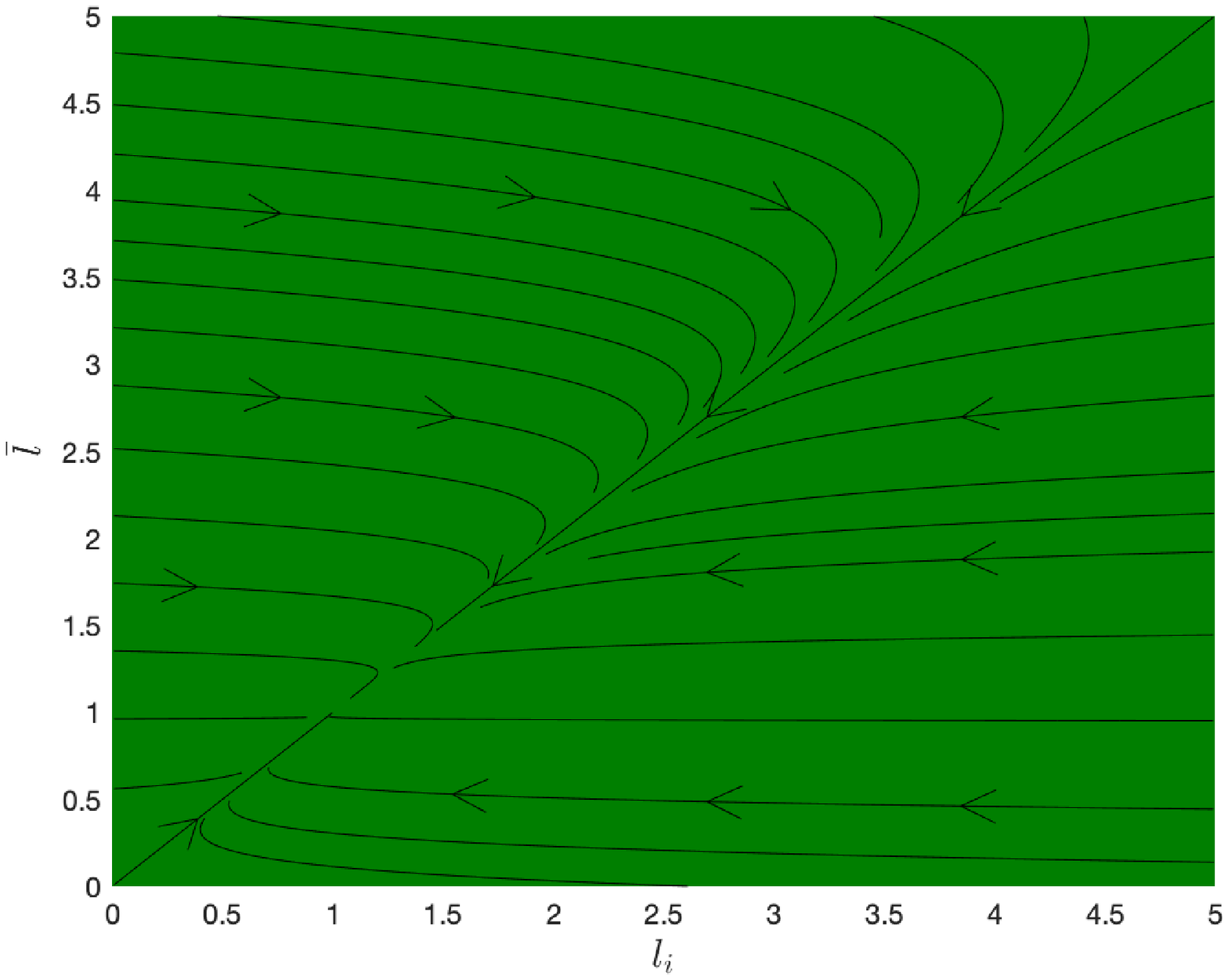}}}\hspace{2pt}
\caption{In this linear system, there is only one equilibrium and this equilibrium is asymptotically stable in large. So we can see that every node in this figure will converge to the only equilibrium when $t\xrightarrow[]{}\infty$. So the ROA is not affected by $w_i$.} 
\label{linear_system}
\end{figure*}

\begin{figure}[tb]
    \centering
    \resizebox*{9cm}{!}{\includegraphics{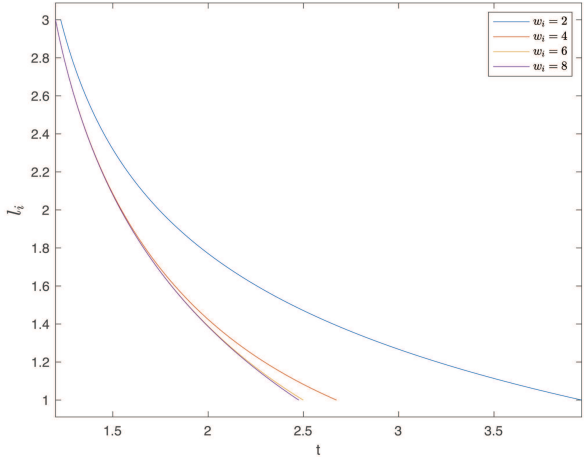}}
    \caption{When different nodes have different network topology $w_i$, the convergence speed of the node from the same initial condition to the equilibrium is different. And we found that speed increase with $w_i$.}
    \label{linear_convergence}
\end{figure}

Since we have compressed the network to a two-dimensional system, we need to use the 2D Hamilton-Jacobi equation 
\begin{equation}
    v_t+H(x, y, t, v_x, v_y) = 0.
\end{equation}

Let $x_i, y_i$ be the $(i, j)$ node of a 2D lattice grid with the uniform spacing $\Delta x$ in x-direction and $\Delta y$ in y-direction. The semi-discrete form of the WENO scheme is

\begin{equation}
    \frac{dv_{i,j}}{dt}=L(v)_{i,j}=-\widehat{H}(x_i,y_i,t,v_{i,j},v^{+}_{x,i,j},v^{-}_{x,i,j},v^{+}_{y,i,j},v^{-}_{y,i,j}),
\end{equation}
where $v_{k,l}=v(x_k,y_l), \Delta_x^{+}=v_{k+1, l}-v_{k,l}, \Delta_y^{+}=v_{k, l+1}-v_{k,l}, \widehat{H}$ is a Lipschitz continuous monotone flux consistent with $H$. The estimation of $\frac{\partial v}{\partial x}$, $\frac{\partial v}{\partial y}$ in 2D Hamilton-Jacobi equation is similar to 1D as shown above, and the details have been explained clearly in \cite{jiang2000weighted}.
Based on the approximation of $\frac{\partial v}{\partial x}$, $\frac{\partial v}{\partial y}$, we can calculate $\frac{\partial v}{\partial t}$ by Runge-Kutta schemes \cite{shu1988total}. The initial value problem can be specified as  
\begin{equation}
\begin{split}
    &\frac{\partial v}{\partial t} = L(v)\\
    &v(x,y,0)= v_0(x, y).
\end{split}
\end{equation}
The fourth order TVD (total variation non-increasing) Runge-Kutta is

\begin{equation}
    \begin{split}
      &v_{(1)}=v_{(0)} + \frac{\Delta t}{2}L(v_{(0)})\\
      &v_{(2)}=v_{(1)} + \frac{\Delta t}{2}(-L(v_{(0)})+L(v_{(1)}))\\
      &v_{(3)}=v_{(2)} + \frac{\Delta t}{2}(-L(v_{(1)})+2L(v_{(2)}))\\
      &v_{(4)}=v_{(3)} + \frac{\Delta t}{6}(L(v_{(0)})+2L(v_{(1)})-4L(v_{(2)})+L(v_{(3)})).
    \end{split}
\end{equation}
The initial solution $v_0(x)$ can be estimated by 
\begin{equation}
    v_0(x,y)=\sqrt{(x-x_e)^2+(y-y_e)^2}-c,
\end{equation}
where $(x_e, y_e)$ is the equilibrium of the 2D network.

\section{Results and Analysis}

In this section, we demonstrate the results of the ROA and analyse the relationship between network topology (local degree) and the ROA.

First, let us investigate the results of linear system. Here we assume that $f(x) = \beta(1-x), g(x) =  \gamma(x)$, where $\beta = 1, \gamma=1$. So, $\dot{l_i}=1-l_i+w_i(\bar{l}-l_i), \bar{l}=1-\bar{l}$. The Hamilton-Jacobi equation is: $v_t + H(l_i,\bar{l}, t, v_{l_i}, v_{\bar{l}})$. The equilibrium is $(1,1)$. So we estimate the initial solution by $v_0(l_i, \bar{l})=\sqrt{(l_i-1)^2+(\bar{l}-1)}-0.1$.
The graph of ROA is shown in Fig.~\ref{linear_system}.

\begin{figure*}[tb]
\centering
\subfigure[ROA of system when $w_i=1$]{%
\resizebox*{6cm}{!}{\includegraphics{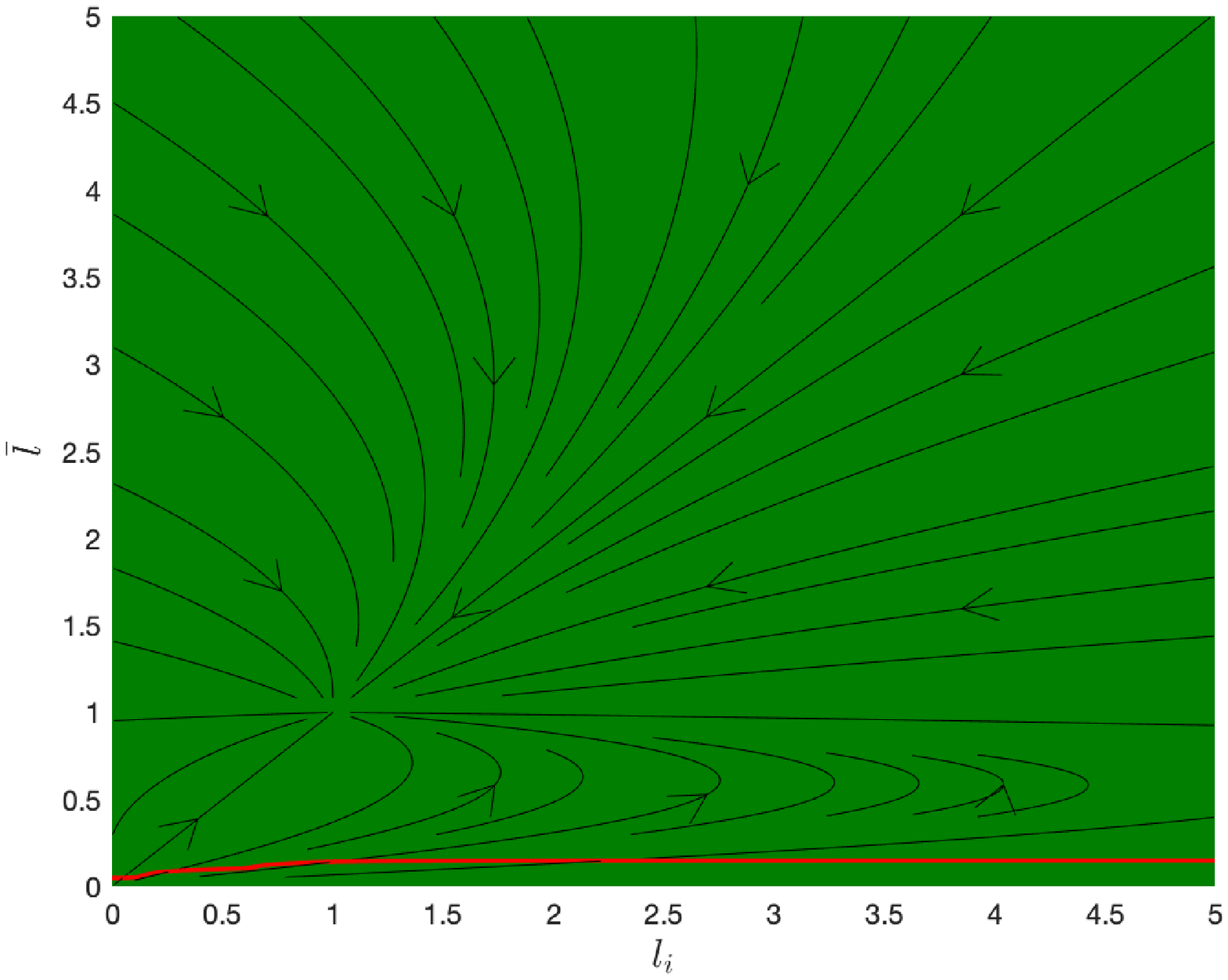}}}\hspace{2pt}
\subfigure[ROA of system when $w_i=3$]{%
\resizebox*{6cm}{!}{\includegraphics{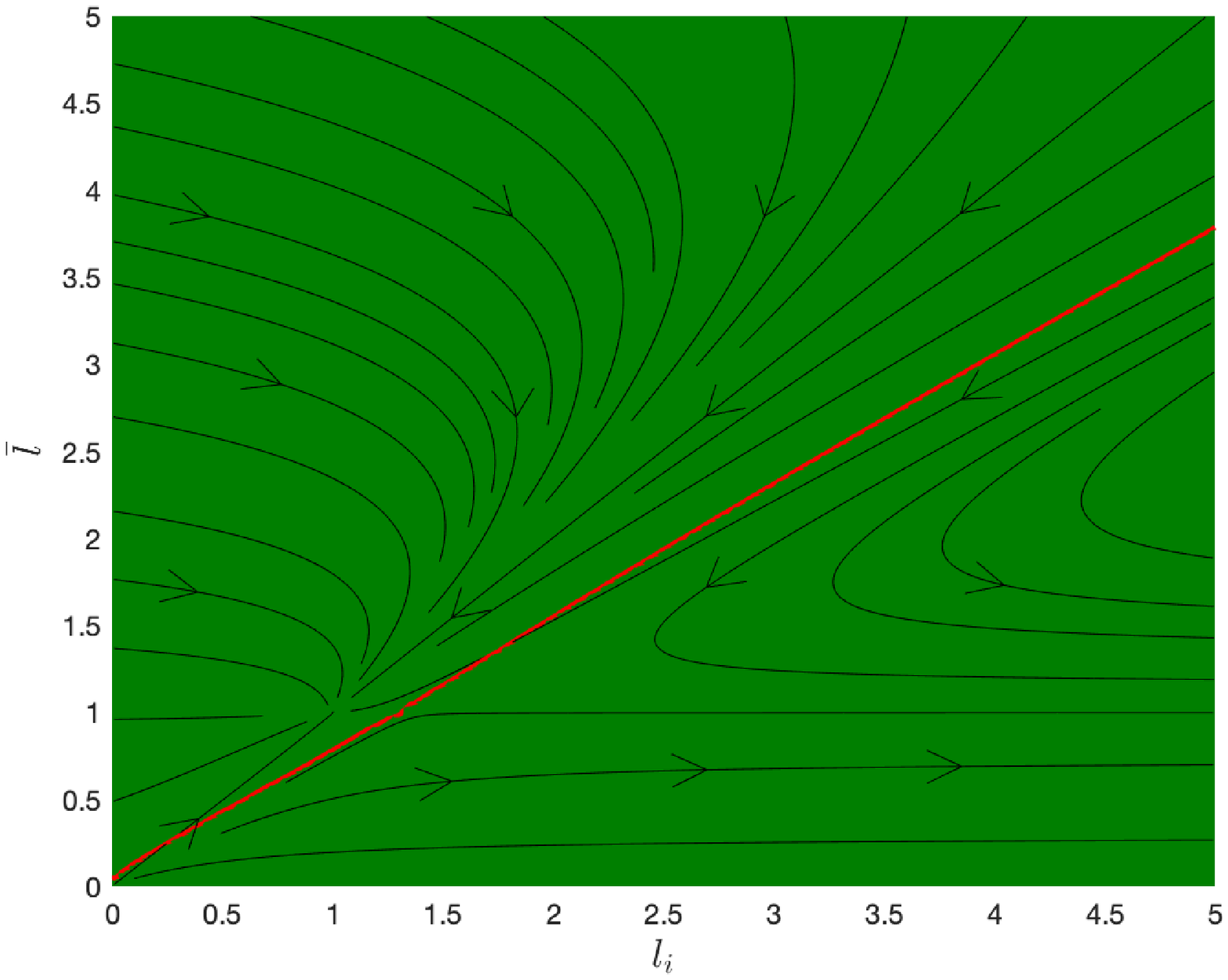}}}\hspace{2pt}
\subfigure[ROA of system when $w_i=5$]{%
\resizebox*{6cm}{!}{\includegraphics{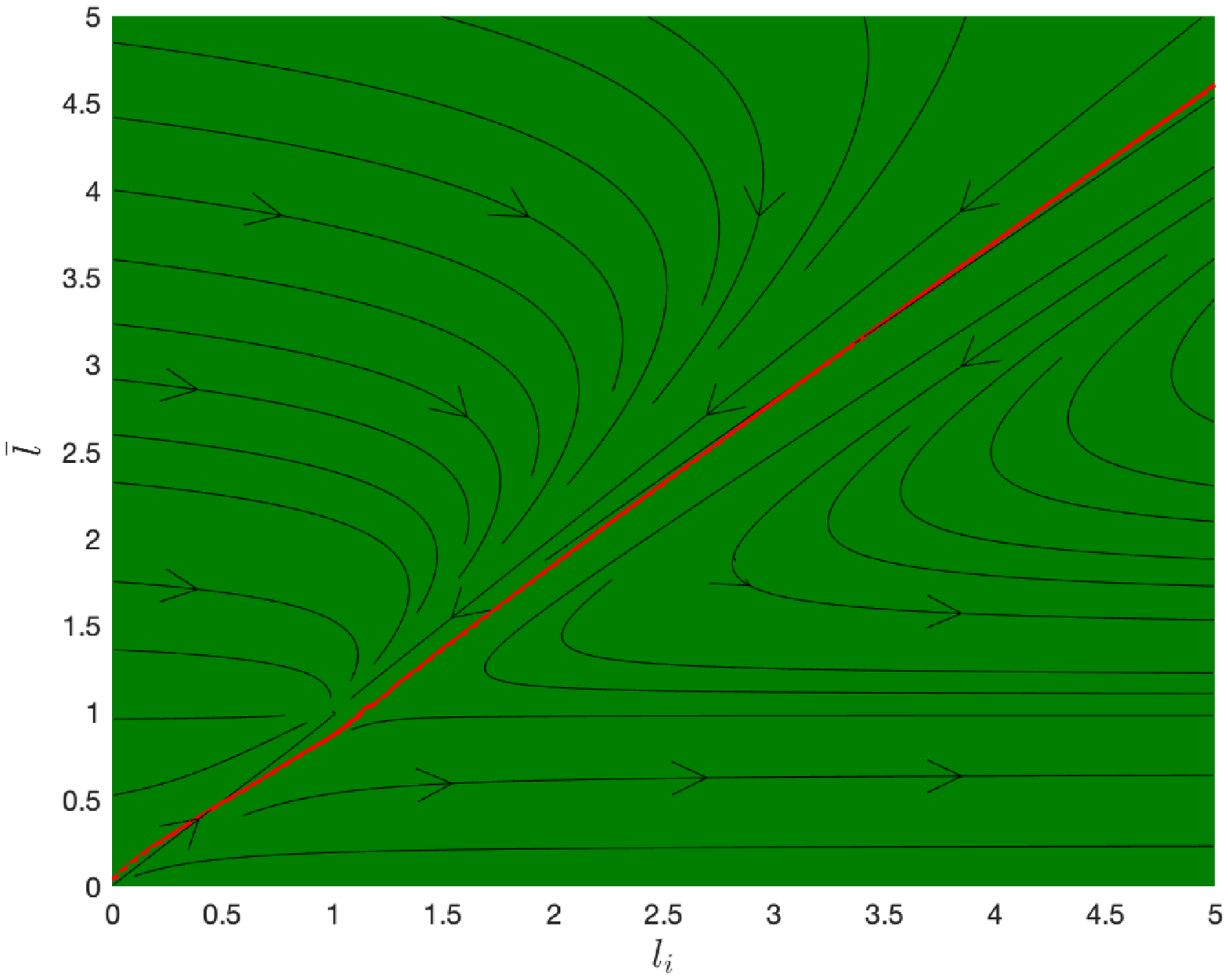}}}\hspace{2pt}
\subfigure[ROA of system when $w_i=7$]{%
\resizebox*{6cm}{!}{\includegraphics{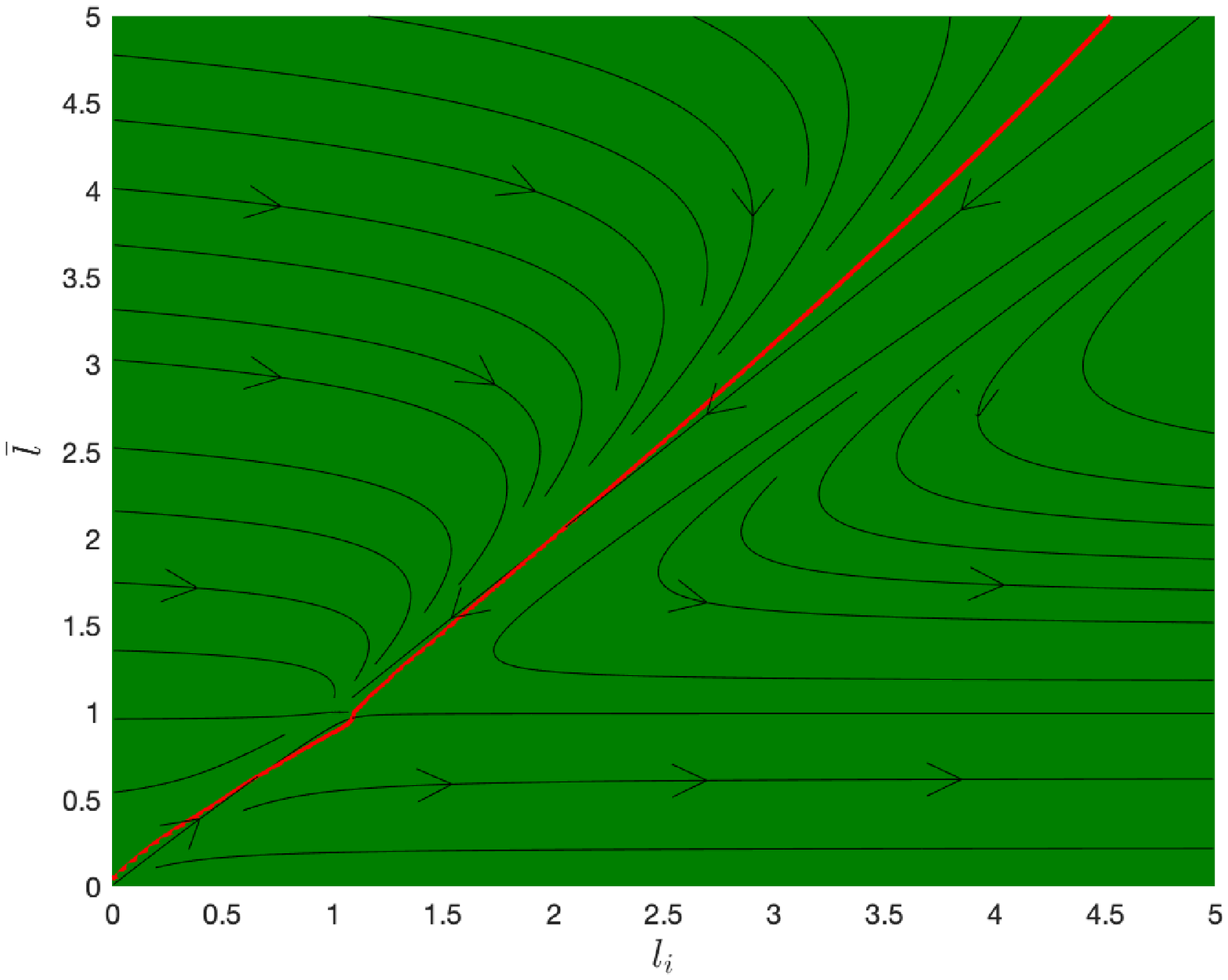}}}\hspace{2pt}
\caption{The four pictures show the estimation of ROA of nonlinear system when $w_i$ are different. It shows that, the value of $w_i$ has an obvious effects on the ROA. With the increase of $w_i$, the ROA decreases.} 
\label{nonlinear_system}
\end{figure*}

\begin{figure}[tb]
    \centering
    \resizebox*{9cm}{!}{\includegraphics{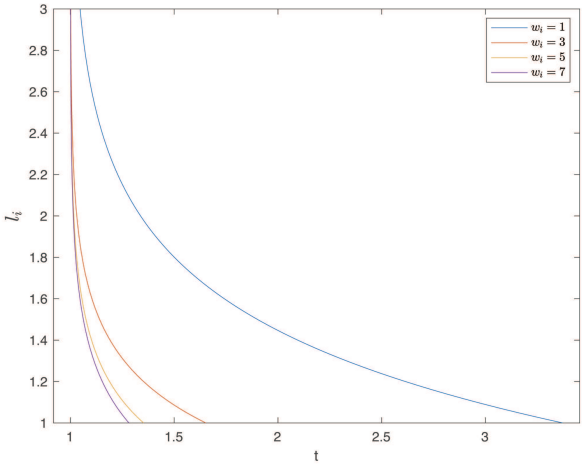}}
    \caption{In nonlinear system. When different nodes have different network topology $w_i$, the cost time of the node from the same initial condition to the equilibrium is different. And we found that speed increase with $w_i$.}
    \label{nonlinear_convergence}
\end{figure}
Since the linear system have only one asymptotically stable equilibrium which have been proved in the aforementioned analysis, all nodes in the system will converge to the equilibrium at last. In Fig~\ref{linear_system}, we can see that when $w_i=2, 4, 6, 8$, nodes in the whole region will converge to the only equilibrium. Therefore, the network topology $w_i$ does not affect the exact ROA in linear system. The only difference for various $w_i$ is the converging trajectory. By solving the system of non-homogeneous linear equations, we find that the network topology (local degree) $w_i$ have an obvious effect on the convergence speed to the equilibrium (shown in Fig.~\ref{linear_convergence}). The increasing of $w_i$ helps the node quickly converge to the equilibrium which is beneficial to the stability of the node as well as the system. This is because $w_i$ characterizes the number of BSs which can offload demand from or to BS $i$. Therefore, a large value of $w_i$ means that BS $i$ is capable of sharing load demand with a lot of neighbour BSs, which is beneficial to converge to the desirable equilibrium.

Alternatively, the analysis and numerical results concerning nonlinear system are provided below. For ease of reading, we assume that $f(x) = x(1-x), g(x) = \gamma_1(x-a)^2+ \gamma_2(x-a)$, where $\beta = 1, \gamma_1=1, \gamma_2=-0.1, a=0$, ($f(x), g(x)$ can have higher order and the method proposed in this paper is also valid.). So, $\dot{l_i}=l_i(1-l_i)+w_i((\bar{l}-l_i)^2-0.1(\bar{l}-l_i)),\dot{\bar{l}}=\bar{l}(1-\bar{l})$. When $w_i$ get different values, the results are shown in Fig.~\ref{nonlinear_system}. From these figures, we can see that with the increase of $w_i$, the ROA of this nonlinear system decreases. This is because, with the rise of $w_i$, another equilibrium will approach the desirable equilibrium $l_i=1$ more closely. Some nodes close to the desirable equilibrium will be attracted to another equilibrium. Hence, the ROA of the desirable equilibrium reduces. Also, we find that the trajectory to equilibrium is different when $w_i$ is various. So, we solve the system of non-homogeneous linear equations to clarify the effects of $w_i$ on the convergence speed to the equilibrium which is shown in Fig.\ref{nonlinear_convergence}. The pictures show that with the rise of $w_i$, the cost time of convergence to the equilibrium reduces. Therefore, in nonlinear system, the network topology $w_i$ affects the BSs' load balance from two aspects: ROA and convergence speed to the equilibrium. The increase of $w_i$ decrease ROA but increases the convergence speed.

% === IV. Transistor Class-F inv Rectifier ========================================
% =================================================================================
\section{Conclusion and Future Work}

In this paper, we proposed an analysis framework to estimate the ROA for traffic load balancing in complex network. We compressed this high dimensional system to a 2D system from the perspective of network theory, and then used the LSM to estimate ROA. Also, we analysed the relationship between ROA and network topology. We found that in linear system, the network topology's local degree $w_i$ would not affect the ROA, but would affect the speed of convergence to the equilibrium. In a non-linear system, network topology have effects on ROA as well as the convergence speed. Therefore, this paper used a LSM technique to estimate the ROA in the complex network system and identified the impact of network topology on controlling its stability. However, we still do not know the effects of network structure on the ROA of each BS. We also believe more work is needed to identify mesoscopic (e.g., community clusters) network structure properties on the stability of the network.

\ifCLASSOPTIONcaptionsoff
  \newpage
\fi

\bibliographystyle{IEEEtran}
\bibliography{Ref}

\vfill


\end{document}